\documentclass[aps,reprint]{revtex4-1}

\usepackage{epsfig,graphics,color}
\usepackage{graphicx}% Include figure files
\usepackage{dcolumn}% Align table columns on decimal point
\usepackage{bm}% bold math

\begin{document}

\title{Uncertainty Principles for Teaching Laboratories}

\author{Kevin L. Haglin}
\email[]{klhaglin@stcloudstate.edu}
%\thanks{}
%\altaffiliation{}
\affiliation{Department of Physics and Astronomy, Saint Cloud State University,
720 Fourth Avenue South, St. Cloud, MN 56301}

\date{\today}

\begin{abstract}
Educators must make decisions about learner expectations and
skills on which to focus when it comes to laboratory activities.  There 
are various approaches but the general pattern is to encourage students to 
measure ordered pairs, plot a graph to establish linear dependence, and then 
compute the slope of the best-fit line for an eventual scientific conclusion.  
To assist educators when they also want to include slope 
uncertainty dependent upon measurement uncertainty as part of the
expected analysis, we demonstrate a
physical approach so that both educators and their 
students have a convenient roadmap to follow.
A popular alternative that educators often choose is to 
rely solely on statistical metrics to establish
the tolerance of the technique, but we argue the statistical
strategy can distract
students away from the true meaning of the 
uncertainty that is inherent in the act of making the measurements.
We will carry these measurement error bars
from their points of origin through the regression analysis to 
consistently establish the physical error bars for the slope and
the intercept.  We then 
demonstrate the technique using an introductory physics experiment 
with a purpose of measuring the speed of sound in air.
\end{abstract}


\maketitle 

\section{Introduction}
\label{intro}

Laboratory activities seem to be nearly sacred to the physics teaching
and learning community.  The experimental aspect of the discipline 
is one of the primary distinguishing features between physics and say
mathematics, or even philosophy.  We are not just finding
mathematical models to describe the physical systems we study, but we
are also confronting the models by going into the laboratory and 
letting nature decide about the quality of the models.  It's what we do!
Important goals of physics experiments oftentimes involve measuring a physical 
quantity of interest or verifying a physical law.  In these cases it 
becomes very important to establish the tolerance of the instrumentation 
used and of the technique employed.  How 
should one report the measured physical quantity at the end of the experiment?
And how does one know if the law is actually verified?  It is fair to say 
that students sometimes have difficulty with these 
questions and even with the meanings behind the terms\cite{pollard,wan,wibig}.  
We will get to
these very important questions in due course. The punch line is that
a complete treatment of measurement uncertainty is critically important
to the endeavor and in fact, the earlier questions cannot be answered 
without it. 

Consistent treatment and expression of uncertainty is 
typically not undertaken in most laboratory activities.  Presumably
it is because the uncertainty techniques are thought to be too burdensome
on the experimenters or perhaps not worth the extra effort.  In fact,
the gold standard on the topic often cited is ``An Introduction to 
Error Analysis'' 
by John. R. Taylor\cite{taylor},
and even there a shortcut is presented that relies solely on statistical
metrics.  We will show
below how straightforward it is to consistently accomplish a very
high level of sophistication in the uncertainty analysis and consequently
avoid the pitfalls and ambiguities of other approaches.  A popular alternative
to completing a full uncertainty analysis is to 
encourage the notion of finding a 
percent error from some expected or predetermined value and using
that metric as a stopping point.  At other times
instructors will encourage the use of a tool called LINEST, 
which establishes
the statistical fluctuations in the data set relative to the
trendline and then models these residuals as a data set exhibiting a 
normal (Gaussian) {\it\/t\/} distribution\cite{linest}. 
If asked, how does 
the software
compute that number? The answer might involve a bit of confused rhetoric.
The linest formulas for the standard deviations of the slope and
intercept
will be articulated below to guide educators through those challenges
and to shed light on the differences between the statistical approach
and physical uncertainty approach.

Equipped with a percent error or an estimate from LINEST for
the statistical reliability of the slope,
students are then asked to make quantitative conclusions based upon these 
metrics.   This is good, but an even better way to proceed is to
establish the actual uncertainty in the slope determined by the physical
processes of making the measurements.  
Consistency ought to be the goal of 
all laboratory teaching and learning activities and must be a strict
requirement for the research arena.
It is singularly important that
educators use proper uncertainty procedures so that 
learners can have the opportunity to develop these high-quality 
experimental skills.

With that goal in mind, we proceed to discuss best-practice
design and implementation for a general experiment.  
Development is most often guided by a
mathematical model which establishes a linear relationship amongst the
accessible variables.  Then a feasible range of values for what 
might be called the
independent variable is tested and corresponding values of a dependent
variable are measured.  The slope of the linear relationship is
either equal to the physical quantity of interest or at worst related
to the physical quantity in some straightforward mathematical way.  That is, 
once the slope is computed, the physical quantity of interest is well
within reach.

Consequently, a typical experimental endeavor is to measure ordered pairs
of data ({\it\/x\/}$_{\it\/i}$, {\it\/y\/}$_{\it\/i}$) and then
plot a graph of {\it\/y\/} versus {\it\/x\/}.
Here we need also the uncertainties associated
with each of these values annotated as 
({\it\/d\/x\/}$_{\it\/i}$, {\it\/d\/y\/}$_{\it\/i}$).  It is tempting
to skip this step for efficiency, but it is imperative that due diligence is
given to establishing these values.
The rule of thumb is that in the absence of other limitations leading to 
the identification of the values, the uncertainty is one half of the 
smallest tick marks ({\it\/i.e.\/} the resolution) of the instrument 
used to make the measurement.  For instance, if a
meter stick is used which has resolution down to the millimeter markings,
then the uncertainty in the measured value would be at minimum
$\pm$ 0.5 mm.
One wonders how these uncertainties propagate throughout the entire analysis
to produce a tolerance for the measured quantity.  The result of the 
experiment must be reported as ``my result $\pm$ $\Delta$(my result)".  
Only then is one prepared to formulate a consistent scientific conclusion.

The paper is organized as follows.  We remind the reader about the
general procedure from multivariable calculus for propagation of 
uncertainties when a (smooth) function
of one or more variables is considered.  This is undertaken
in section~\ref{computedfofxyz} and results in
the so-called quadrature formula.
Then we review the least-squares method and recall the formulas
for the slope and intercept of the best-fit line in 
section~\ref{dalphabeta}.  The next
task is to derive formulas for the uncertainties in the slope and
intercept, which are also presented in section~\ref{dalphabeta}.  
These formulas are what we are referring to as the 
``Uncertainty Principles'' for teaching laboratories.
Discussion follows in section~{\ref{linestapproach} of a more commonly 
used approach which establishes ``statistical variations" in the slope
and intercept.  
A statistical analysis emphatically does not take the place of a 
consistent uncertainty
analysis~\cite{dogma}, but instead can be used as a comparison
and to identify when outliers within
the data set are introducing extra challenges.
Then, in section~\ref{soundspeed}
we demonstrate the technique by carrying out
an experiment to measure the speed of sound in air and
especially to consistently establish the uncertainty in that measured quantity.
A conclusion follows in section~\ref{summary}
to summarize the main findings and to encourage all experimenters
to regularly deploy these techniques.

\section{Uncertainty: General Considerations}
\label{computedfofxyz}

The place to begin the discussion of uncertainty is to 
demonstrate how to compute the uncertainty in a value which is
not accessible directly but is instead 
dependent upon one or more accessible variables.  That is, suppose we have a 
function {\it\/f\/}({\it\/x},
{\it\/y}, {\it\/z}) for which we have measured the variables on which
{\it\/f\/} depends as well
as the uncertainties in those quantities. This means we have used some 
instrument to
produce {\it\/x\/} $\pm$ {\it\/d\/x\/}, 
{\it\/y\/} $\pm$ {\it\/d\/y\/}, 
and {\it\/z\/} $\pm$ {\it\/d\/z\/}.   We need to establish the uncertainty
in the function's value at the measured set of values---and naturally 
we turn to calculus for this.  

The square of the total uncertainty in 
the function due to uncertainties
in the variables upon which it depends is the following (using
the chain rule and assuming independence of 
the variables)\cite{swokowski}
\begin{eqnarray}
\mbox{  \,(\,}\mbox{\it\/d\,f\/}\mbox{\,)}^{\mbox{\,2\/}}
& \mbox{\ =\ } & 
\left({\partial\,\mbox{\it 
f\/}\over\partial\mbox{\it x\/}}\,\mbox{\it d\/x\/}\right)^{\mbox{\/2}}
\mbox{ + }
\left({\partial\,\mbox{\it 
f\/}\over\partial\mbox{\it y\/}}\,\mbox{\it d\/y\/}\right)^{\mbox{\/2}}
\mbox{ + }
\left({\partial\,\mbox{\it 
f\/}\over\partial\mbox{\it z\/}}\,\mbox{\it d\/z\/}\right)^{\mbox{\/2}}
\mbox{\ .}
\nonumber
\end{eqnarray}
A good tactic for visualizing and describing this formula is that
it can be thought of as ``the distance formula in multidimensional 
data space''.
Uncertainties in the variables propagate
through to inform the final uncertainty {\it\/df} in this manner.

As a side note, the uncertainties in each of the variables 
will include their ``net uncertainties''.  By that, we mean everything:
systematic uncertainty (which includes such things as
known instrument bias effects, offset effects,
and other non-random effects), random uncertainty (which includes
measurement uncertainty and other uncontrolled influences exhibiting
random features) and  possibly others~\cite{commentoneverything}.  
This list of sources introduces challenges but also highlights the 
rewards of doing 
experimental work.  In this paper we focus exclusively on measurement 
uncertainty but
stress that other known effects could be included as needed.
Coming back to result of the quadrature formula,
we ought to conclude at this point that 
the ``true'' value of the function is expected to lie somewhere 
within the range
({\it\/f\/} $-$ {\it\/df\/}, {\it\/f\/} $+$ {\it\/df\/}) at the
point ({\it\/x}, {\it\/y}, {\it\/z}).

We are now prepared to carry this formalism forward to data
analysis tasks one often encounters in the laboratory.

\section{Uncertainty: Slope and Intercept}
\label{dalphabeta}

Quantitative experimentation mainly involves
measuring ordered pairs of data with an expectation
that the data are linearly related to one another.  That is, we expect that if
we plot the {\it x\/}$_{\mbox{\it\/i\/}}$'s on the horizontal
axis and the 
{\it y\/}$_{\mbox{\it\/i\/}}$'s on the vertical, there will be a 
linear trend along which the data will follow.  The slope
of the best-fit line $\Delta${\it y\/}/$\Delta${\it x\/} is typically 
{\it\/the\/}
meaningful, and sought-after quantity.  In modern times we rely
solely on pre-packaged software\cite{analysissoftware} to 
compute the slope and intercept for the trendline of the data set.  

However, in order to derive the uncertainty in the slope (and intercept) 
we must first
establish closed-form expressions for the slope and intercept from 
linear regression analysis and then 
apply the quadrature formula to them.
There are alternate approaches to curve 
fitting and linearization with arguably more 
sophistication\cite{matthewswalker,richards},
but
for our purposes here, this is sufficient.
The goal is the following.  Given a set of ordered pairs
({\it x\/}$_{1}$,{\it y\/}$_{1}$),
({\it x\/}$_{2}$,{\it y\/}$_{2}$),
\ldots,
({\it x\/}$_{N}$,{\it y\/}$_{N}$),
we seek the equation of the line 
\begin{eqnarray}
	{\it y\/} & = & \alpha{\it\,x\/} + \beta 
\end{eqnarray}
which best describes the data.  It boils down to
searching for the  slope and the intercept of the line, namely $\alpha$
and $\beta$\cite{freundwalpole1}.

We begin by defining $Q$, the sum of the squares of the
differences from the line to each data point (sum of squares of
the lengths of the dotted vertical line segments in Fig.~\ref{lr},
sometimes referred to as the residual values)
\begin{eqnarray}
{Q}\,{\hskip 0.1 em}\mbox{\rm(}\alpha\mbox{\rm\,,}\,\beta\mbox{\rm)} 
& \equiv & \sum_{\mbox{\it\,i\rm\,=\,1}}^{\mbox{\it\,N\/}}
\left(\alpha\mbox{\it\,x\/}_{\mbox{\it\,i}}
\mbox{\rm \ + \ } \beta 
\mbox{\rm \ -- \,\it\,y\/}_{\mbox{\it\,i\/}}\right)^{\mbox{\rm\,2}}
\mbox{\rm .}
\nonumber
\end{eqnarray}

\begin{figure}[!h]
\begin{center}
\includegraphics[scale=0.45]{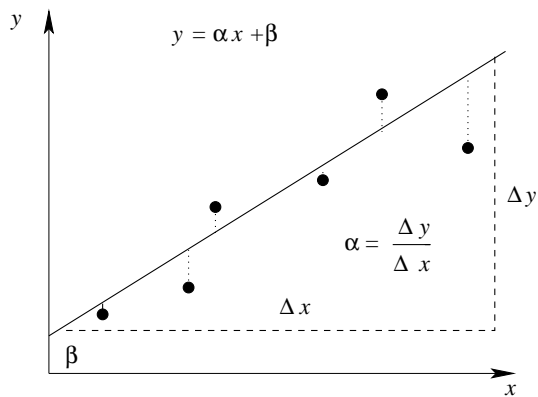}
	\caption{A generic data set shown as well as the
	linear regression line.  Squares of the vertical
	distances from 
	data to the line contribute to the $Q$ value.}
\label{lr}
\end{center}
\end{figure}

\noindent Certainly the best-fit line ought to be the one which
minimizes this quantity $Q$.   We accomplish 
this minimization
by taking partial derivatives with respect to $\alpha$ and $\beta$,
and separately setting them to zero.

\begin{eqnarray}
{\partial{Q}\over\partial\alpha} & \mbox{\rm = } & 
\mbox{\rm 0} \quad
\mbox{\rm = } \quad
\sum_{\mbox{\it\,i\rm\,=\,1}}^{\mbox{\it\,N\/}}\mbox{\rm\,2}\,\left(
\alpha\mbox{\it\,x\/}_{\mbox{\it\,i}}\mbox{\rm\ + \ }\beta
\mbox{\rm \ --\ \it y\/}_{\mbox{\it\,i}}
 \right)\mbox{\it\,x\/}_{\mbox{\it\,i}}
\nonumber\\
{\partial{Q}\over\partial\beta} & \mbox{\rm = } & 
\mbox{\rm 0} \quad 
\mbox{\rm = } \quad
\sum_{\mbox{\it\,i\rm\,=\,1}}^{\mbox{\it\,N\/}}\mbox{\rm\,2}\,\left(
\alpha\mbox{\it\,x\/}_{\mbox{\it\,i}}\mbox{\rm\ + \ }\beta
\mbox{\rm \ --\ \it y\/}_{\mbox{\it\,i}}
\right)
\mbox{\rm .}
\nonumber
\end{eqnarray}
We can simplify the equations using the definitions of averages,
\begin{eqnarray}
\langle\mbox{\it\/x\/}\rangle & \equiv & {\mbox{\rm\,1\/}\over\mbox{\it\,N\/}}
\sum_{\mbox{\it\,i\rm\,=\,1}}^{\mbox{\it\,N}}\mbox{\it\,x\/}_{\mbox{\it\,i\/}}
\nonumber\\
\langle\mbox{\it\/y\/}\rangle & \equiv & {\mbox{\rm\,1\/}\over\mbox{\it\,N\/}}
\sum_{\mbox{\it\,i\rm\,=\,1}}^{\mbox{\it\,N}}\mbox{\it\,y\/}_{\mbox{\it\,i\/}}
\mbox{\rm,}
\nonumber
\end{eqnarray}
to write down the following two equations
\begin{eqnarray}
\alpha\sum_{\mbox{\it\,i\rm\,=\,1}}^{\mbox{\it\,N}}
\mbox{\it\,x\/}_{\mbox{\it\,i\/}}^{\mbox{\rm\,2\/}}
 \mbox{\rm \ + \ }
\beta\,\mbox{\it N\/}\langle\mbox{\it\/x\/}\rangle
 \mbox{\rm \ -- \ }
\sum_{\mbox{\it\,i\rm\,=\,1}}^{\mbox{\it\,N}}
\mbox{\it\,x\/}_{\mbox{\it\,i\/}}
\mbox{\it\,y\/}_{\mbox{\it\,i\/}}
& \mbox{\rm = }  & \mbox{\rm 0}
\nonumber\\
\alpha\mbox{\it\,N\/}\langle\mbox{\it\/x\/}\rangle 
 \mbox{\rm \ + \ }
\beta\mbox{\it\,N\/}
 \mbox{\rm \ -- \ }
\mbox{\it\,N\/}\langle\mbox{\it\/y\/}\rangle
& \mbox{\rm = }  & \mbox{\rm 0}
\mbox{\rm\,.}
\nonumber
\end{eqnarray}
These can be readily solved for $\alpha$ and $\beta$.  The results 
are
\begin{eqnarray}
\alpha & \mbox{\rm\ =\ } & 
	{\left(
\sum_{\mbox{\it\,i\rm\,=\,1}}^{\mbox{\it\,N\/}}
\mbox{\it\,x\/}_{\mbox{\it\,i\/}}
\mbox{\it\,y\/}_{\mbox{\it\,i\/}}
\mbox{\rm \ -- \ } \mbox{\it\,N\/}\langle\mbox{\it\/x\/}\rangle
%\langle\mbox{\it\/y\/}\rangle\right)\left/
\langle\mbox{\it\/y\/}\rangle\right)\over
\left(\sum_{\mbox{\it\,i\rm\,=\,1}}^{\mbox{\it\,N\/}}
\mbox{\it\,x\/}_{\mbox{\it\,i\/}}^{\mbox{\rm\,2}}
\mbox{\rm \ -- \ } \mbox{\it\,N\/}\langle\mbox{\it\/x\/}
	\rangle^{\mbox{\rm\,2}}\right)},
%	\rangle^{\mbox{\rm\,2}}\right)\right.,
\nonumber
\end{eqnarray}
(which can be more compactly written as)
\begin{eqnarray}
	\alpha
	& \ \mbox{\ =\ } & { \langle{x\,y\/}\rangle \ - \ 
	\langle{x}\rangle\langle{y}\rangle \over 
	\langle{x}^{2}\rangle \ - \ \langle{x}\rangle^{2}  } 
\label{slope}
\end{eqnarray}
and
\begin{eqnarray}
\beta & \mbox{\rm\ =\ } & \langle\mbox{\it\/y\/}\rangle \mbox{\rm \ -- \ }
\alpha\langle\mbox{\it\/x\/}\rangle
\mbox{\rm .}
\label{intercept}
\end{eqnarray}
The formulas for slope and intercept are of course well known but
restated here for completeness and to facilitate quadrature analysis
below.

It is typically the case in experimental endeavors that the slope of 
the best-fit line represents ``the'' best estimate for a particular 
physical quantity.  Significant progress has been made at this point.
However, since our main purpose here is to express uncertainties in 
the slope and
intercept, we move onward---guided by the quadrature formula from
Sect.~\ref{computedfofxyz}.  
The slope and intercept are now functions of many numbers
{\it\/x\/}$_{\it\/i}$
and {\it\/y\/}$_{\it\/i}$,
but the quadrature formula is still the appropriate tool to 
establish net measurement uncertainties. 

\subsection{Uncertainty in the Slope}

Using the quadrature formula we get the uncertainty in the slope to
be
\begin{widetext}
\begin{eqnarray}
	d\alpha & = & \sqrt{
	\sum_{i=1}^{N}\left[
\left({\partial\alpha\over\partial\/x_{i}}\right)^{2}(d\,x_{\it\/i})^{2} 
	\mbox{\, + \,}
\left({\partial\alpha\over\partial\/y_{i}}\right)^{2}(d\,y_{\it\/i})^{2} 
		\right]\  }
\label{dslope}
\end{eqnarray}
where
\begin{eqnarray}
	{\partial\alpha\over\partial\/x_{i}} & = & 
	{1\over{N}}\left({y_{i}-\langle{y}\rangle}\over{\langle{x}^{2}\rangle
	\ - \ \langle{x}\rangle\/^{2}}\right)
	\ - \
	{2\over{N}}\left(\left(x_{i}\ - \ \langle{x}\rangle\right)\cdot
	\left(\langle{x\,y}\rangle \ - \ 
	\langle{x}\rangle\langle{y}\rangle\right)\over\left[ 
	\langle{x}^{2}\rangle \ - \
	\langle{x}\rangle\/^{2}
	\right]^{2}\right)\ ,
\end{eqnarray}
\end{widetext}
and
\begin{eqnarray}
	{\partial\alpha\over\partial\/y_{i}} & = & 
	{1\over{N}}\left({x_{i}-\langle{x}\rangle}\over{\langle{x}^{2}\rangle
	\ - \ \langle{x}\rangle\/^{2}}\right).
\end{eqnarray}

We stress at this point that the uncertainty in the slope is determined
by the overall set of uncertainty values established previously. Furthermore,
the uncertainty in the slope increases or decreases as the individual 
measurement uncertainties dictate.  This is what was meant earlier by
the phrase ``physical uncertainty''.   

\subsection{Uncertainty in the Intercept}

Applying the quadrature formula to the regression line's {\it\/y\/} intercept,
we get the uncertainty in the intercept to be
\begin{widetext}
\begin{eqnarray}
	d\beta & = & \sqrt{
	\sum_{i=1}^{N}\left[
\left({\partial\beta\over\partial\/x_{i}}\right)^{2}(d\,x_{\it\/i})^{2} 
	\mbox{\, + \,}
\left({\partial\beta\over\partial\/y_{i}}\right)^{2}(d\,y_{\it\/i})^{2} 
         \right]
	\mbox{\, + \,}
\left({\partial\beta\over\partial\alpha}\right)^{2}(d\alpha)^{2}\ 
	                 }
\label{dintercept}
\end{eqnarray}
\end{widetext}
where
\begin{eqnarray}
	{\partial\beta\over\partial{x}_{i}} & \ = \ &
	\ - \
	{\alpha\over{N}} \ - \ 
	\langle{x}\rangle{\partial\alpha\over\partial{x}_{i}},
\end{eqnarray}
\begin{eqnarray}
	{\partial\beta\over\partial{y}_{i}} & \ = \ &
	 {1\over{N}} \ - \ \langle{x}\rangle{\partial\alpha
\over\partial{y}_{i}},
\end{eqnarray}
and lastly,
\begin{eqnarray}
	{\partial\beta\over\partial\alpha} & \ = \ & -\,\langle{x}\rangle.
\end{eqnarray}
We are now ready to compute the slope and its uncertainty 
from Eqs.~({\ref{slope}}) and ({\ref{dslope}}) as well as the intercept
and its uncertainty from Eqs.~({\ref{intercept}}) and ({\ref{dintercept}}).
It is very important to reiterate that the uncertainty in the slope and the 
uncertainty in the intercept are driven solely by uncertainties in the
measurements themselves.  That means these metrics are directly
connected through physical activities to the experimental 
technique.  In contrast, statistical variation metrics 
are dependent upon quite different aspects of the data set.
Statistical metrics will be discussed in the next section.

\section{Standard Deviations: Slope and Intercept}
\label{linestapproach}

Admittedly, carrying out consistent uncertainty analysis seems a 
bit laborious.  Consequently, it is often the case that students are 
asked to use a much simpler alternative tool to merely approximate 
the uncertainty.  That 
is, students are often asked to use Excel
to compute the statistical variation in the data and its
influence on the slope---the so-called
standard deviation of the slope.   
This is done using the ``LINEST" function.
We must be careful however, not to confuse the two approaches as
being interchangeable or equally meaningful.  They are not!  Indeed, 
they have drastically 
different meanings and carry with them very different information.

The linest function in Excel is exceedingly convenient to use for
establishing statistical metrics, but it cannot express
uncertainties\cite{statutility}
Furthermore, one cannot look into the code to get at the mathematical
formulas for these statistical metrics leaving the user wondering how
it was computed.  In order to assist instructors to correctly
respond when students ask how does Excel compute the standard
deviation of the slope and intercept?, we include here the mathematical
formulas that Excel uses to compute them.

First off, assumptions are made about statistical independence of 
measurements within the data set.  A random variable {\it\/z\/} is 
constructed within this approach exhibiting a standard normal distribution 
(a reasonable
assumption for most experimental endeavors).  From here a random variable 
{\it\/t} is created using various moments of the distribution and
related statistical features within the set of measurements relative
to the trendline.  Specific details
are not critical to this discussion, but the interested reader is encouraged
to consult the literature~\cite{freundwalpole2}. The new random 
variable has a {\it\/t\/} distribution with
({\it\/N\/} -- 2) degrees of freedom.  This is why the title of the 
function in Excel is called linest, because it is based upon the 
line's ``{\it\,t\,}'' statistics---dubbed the ``LINEST"
function.

The standard deviation of the slope computed within
LINEST comes from the following formula~\cite{freundwalpole2}

\begin{eqnarray}
	\sigma_{\alpha} & = & \sqrt{\sigma^{2}\over\,S_{xx}\ }
	\nonumber\\
	\nonumber\\
	& = & \sqrt{{\sum_{i=1}^{N}
	[y_{i}-(\alpha\,x_{i} + \beta)]^2\over(N\,-\,2)
	[N\langle{x}^{2}\rangle - N\langle{x}\rangle^{2}]}}\ .
	\label{linestdslope}
\end{eqnarray}
Typical statistics nomenclature is adopted here in the first
expression, while the follow-up line
fills in some of the practical details\cite{s2andsxx}.
Furthermore, the standard deviation of the intercept
value computed within LINEST comes from\cite{freundwalpole2}

\begin{eqnarray}
	\sigma_{\beta} & = &  \sqrt{{(S_{xx}+N\langle{x}\rangle^{2})\sigma^2
	\over N\/S_{xx}}}\ ,
	\nonumber\\
	\nonumber\\
	& \ = \ & \sqrt{{{\langle{x}^{2}\rangle\left(\sum_{i=1}^{N}[y_{i}-(
	\alpha\,x_{i} + \beta)]^{2}\right)\over[N-2]N(\langle{x}^{2}\rangle
	\ - \ \langle{x}\rangle^{2})}}}\ ,
	\nonumber\\
	\nonumber\\
	& \ = \ & \sigma_{\alpha}\,\sqrt{\langle{x}^{2}\rangle}\ . 
	\label{linestdintercept}
\end{eqnarray}
These formulas serve as the basis for the statistical uncertainties 
in the slope and the intercept produced as
output from the linest function in Excel.  A side note is in order.
As the number of measurements increases without bound, one can see that the
statistical variations in the slope and intercept approach zero
as $\sim 1/\sqrt{N}$, consistent with various expectations.

Let us check in on the uncertainty
expressions for slope and intercept from the quadrature approach.  
Summations are essentially proportional to $N$,
and so one can observe that both equations (\ref{dslope}) and
(\ref{dintercept}) behave in the same way.  That is, they both approach
zero like 1/$\sqrt{N}$ as well, albeit with different pre-multipliers
owing to the presence of measurement uncertainties. 
This is not unexpected since the measurement uncertainty 
stems from a random process
and consequently must have the same large $N$ behavior as does the 
standard deviation.  It is reassuring to note that taking
more and more measurements will not cause the uncertainties in the
slope and intercept to increase but instead tend toward smaller
values.

If it is not feasible to complete a full-blown uncertainty analysis by 
identifying measurement uncertainties for each data point taken and 
propagating those to inform the uncertainties in the slope and intercept, 
then the standard deviations of the slope and intercept are suitable 
alternatives.  
But a competition is not what is intended.  Instead, the statistical
metrics are often very useful to be used in conjunction with the
measurement uncertainty techniques to interrogate data sets. Both 
strategies are encouraged.

\section{Example: Speed of Sound in Air}
\label{soundspeed}
At long last we are prepared to deploy this formalism in an
example from introductory physics endeavors.
It will serve to illustrate how the uncertainty analysis is consistently
applied in a typical experimental setting.
The scientific purpose of this particular experimental activity is to 
measure the speed of sound in air using tuning forks and sound
tubes\cite{klhsoundlab}.  Resonance conditions are readily achieved by adjusting
the length of the columns of air in the sound tubes so that nodes
of the standing sound waves can be recognized and located.  Locations 
of nodes
inform the experimenter about the wavelength (wavelength is twice
the average node-to-node distance).  If one has frequency
and wavelength for a discrete set of frequencies, one can use
$\lambda$\/{\it\/f} = {\it\/v}, or
\begin{eqnarray}
	\lambda & = & {v}\left({1\over\/f}\right)
\end{eqnarray}
to investigate how the wavelength of the standing waves depends upon
the independent variable of inverse frequency.  Then
a plot of $\lambda$ vs. {1/{\it\/f\/}} produces linear
behavior with a slope equal to the speed of sound in air.  But what
is the uncertainty of the technique?

The data displayed in Table~\ref{sounddata} were gathered at 
the author's home institution.
From the tuning fork frequency, we compute the column of inverse frequency. 
Further, using the quadrature formula, we identify
the uncertainty in the inverse frequency as
\begin{eqnarray}
	d(1/f) & = & \sqrt{(-df/f^2)^2} \quad =\quad \left| 
	-df/f^2 \right|.
\end{eqnarray}
Next we comment on the 
uncertainty in the wavelength measurements. Since it becomes somewhat 
difficult to 
say where ``best resonance" conditions persist ({\it\/i.e.} loudest 
sounds) using
only unaided human hearing, there 
seems to be a window
of node locations perhaps as wide as roughly 1.0 cm or more
(meaning {\it\/d\/}$\lambda$
= 0.5 cm) where the true location resides.  See the table for 
specific {\it\/d\/}$\lambda$ values 
established for this activity. 
Measured values in the table carry the available significant figures 
(reportable digits)
while computed columns carry extra digits (a strategy which is consistent
with preferred experimental techniques and expectations). 

\begin{table}
\caption{Frequency and wavelength data are presented here,
	including best-guess uncertainties.}
%Refs.~\protect\cite{jpsixsect2,kevin,linko}.}
	\null
\label{sounddata}
\begin{tabular}{c|c|c|c|c|c}
	{\it\/f} (Hz) & {\it\/df} (Hz)
	& 1/{\it\/}f (sec) & {\it\/d}(1/{\it\/f}) (sec) & $\lambda$ (m) & d$\lambda$ (m) \\
\tableline
	1000. & 1.0 & 1.000E-03 & 1.00E-06 & 0.3432 & 0.005 \\ \hline
	512 & 0.5 &1.953E-03 & 1.91E-06 & 0.661 & 0.008 \\ \hline
	480. & 0.5 & 2.083E-03 & 2.17E-06 & 0.715 & 0.010 \\ \hline
	426.7 & 0.4 & 2.344E-03 & 2.2E-06 & 0.796 & 0.010 \\ \hline
\end{tabular}
\end{table}

\begin{figure}[!h]
\begin{center}
\includegraphics[scale=0.45]{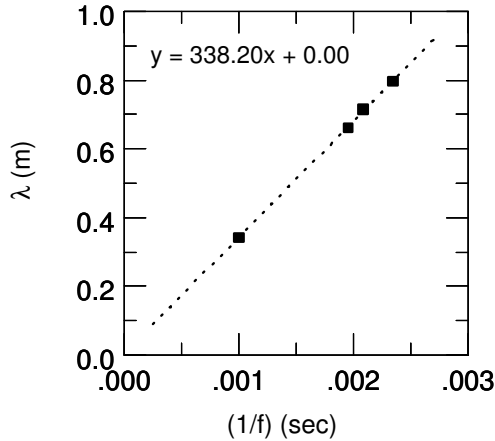}
	\caption{$\lambda$ vs. period for data set in 
	Table~\ref{sounddata}.  
	The best-fit line is dotted and its equation is displayed.}
\label{lvst}
\end{center}
\end{figure}
The data set is plotted in Fig.~{\ref{lvst}.
With these measurements and corresponding uncertainties, the slope and
intercept of the linear regression ({\it\/i.e.} trend-) line can be computed.
Using Eq.~(\ref{slope}) we get

\begin{eqnarray}
	\mbox{\rm\/slope} & = & \mbox{\rm\/338.2048\, m/s}
\end{eqnarray}
and Eq.(~\ref{intercept}) we get
\begin{eqnarray}
	\mbox{\rm\/intercept} & = & \mbox{\rm\/0.00204\, m}.
\end{eqnarray}
The trendline equation on the graph has been formatted to include
up to the hundredths place, and not beyond.  

The uncertainties in these quantities from Eqs.~(\ref{dslope})
and ~(\ref{dintercept}) are computed to be
\begin{eqnarray}
	\mbox{d(\rm\/slope)} & = & \mbox{\rm\/6.7983\, m/s}
\end{eqnarray}
and 
\begin{eqnarray}
	\mbox{d(\rm\/intercept)} & = & \mbox{\rm\/0.01672\, m}.
\end{eqnarray}

For completeness, we include also the results from linest for
the statistical variations (also
calculable using Eqs.~\ref{linestdslope}) and Eqs.~\ref{linestdintercept})).
They are found to be
\begin{eqnarray}
	\sigma_{slope} & = & \mbox{\rm\/5.0420\, m/s}
\end{eqnarray}
for the slope's standard deviation, and 
\begin{eqnarray}
	\sigma_{intercept}
	& = & \mbox{\rm\/0.00965\, m}
\end{eqnarray}
for the intercept's standard deviation. If the statistical variations
in the slope or intercept were significantly greater than the
uncertainties in those quantities expressed earlier, it would 
signal ``outlier challenges'' 
to be addressed separately.  Here we do not have such challenges
as the data are exceptionally linear.

We are now prepared to answer the question:  Can we 
confirm that our 
measurement is consistent with the expected
(accepted) value? We shall respond in the conclusion.

\section{Conclusion}
\label{summary}
Encouraging educators to elevate student
expectations for laboratory activities to include consistent
uncertainty treatments and equipping them with the tools to do so
is the primary purpose of this paper.  If mentors 
do not correctly undertake uncertainty analysis, how does one expect the
students to know what to do?  Let alone, do it!  This is the reason for
the urgency to 
derive the uncertainty formulas for the slope and intercept
within the scaffolding of linear regression analysis 
applied to a general
data set and make those formulas readily available to the community.  These
formulas represent the original contributions of this paper.
The main message of the paper might be this:
The slope of a best-fit line tells a certain story, but
without the physical uncertainty in that value, the scientific 
story cannot be fully told.  And, a mere statistical estimate of the
value is not as desirable as having the uncertainty in the
slope determined by the measurements and their uncertainties.
Moreover, one must report the slope $\pm$ the slope's uncertainty to be 
scientifically honest and complete. If either one of those numbers 
is missing, the story remains unfinished.

Formulas for the slope and the uncertainty in the slope are
derived and included as Eqs.(\ref{slope}) and (\ref{dslope}). 
If the physical quantity of interest in one's experiment requires
the intercept as input as well, then the uncertainty in the intercept is
required to properly report the uncertainty in the physical quantity.
For that reason, derivations of the intercept and its uncertainty are also
included in this work as
Eqs.(\ref{intercept}) and (\ref{dintercept}).  For extra completeness, we also
include the formulas that Excel utilizes within its LINEST function
for the statistical variations (standard deviations)
in the slope and intercept.  These statistical variation metrics
are included in Eqs.(\ref{linestdslope}) and (\ref{linestdintercept}).

Finally, the most complete and consistent method of reporting a
measurement at the end of a technical report is available using these 
techniques.  It will have
the following structure:
\begin{widetext}
\begin{eqnarray}
	\mbox{\rm physical quantity measured} 
	& = & \left[\mbox{\rm my measurement} \pm
	\mbox{\rm d(my measurement)}\right] 
	\mbox{\rm\/(units of physical quantity)}.
\end{eqnarray}
\end{widetext}
We would be remiss if we did not point out that the uncertainties
cannot be reported to a decimal place beyond that which the actual
measured quantity is reportable.  This is the challenge of keeping 
significant figures under control and consistent.
Specifically applied to the sample laboratory activity of measuring
the speed of sound in air it would read
%\begin{widetext}
\begin{eqnarray}
	\mbox{\it\/v}_{sound\ in\ air\/} 
	& = & \left[\mbox{\rm 338} \pm
	\mbox{\rm 7}\,\right] \mbox{\rm (m/s)}.
\end{eqnarray}
%\end{widetext}
If the so-called accepted value of a physical quantity lies within
the range of numbers (measured value $\pm$ uncertainty), then one
can conclude that one has confirmed the accepted value.  For the speed
of sound experiment we have a temperature dependent accepted value to
which we might compare~\cite{{youngfreedman},{serwayvuille}}.  It 
is ({\it\/T} = 22.5$^{\circ}$ C in the room
at the time data were gathered),
\begin{eqnarray}
v & = & {\rm\/331\ m/s}\,\sqrt{1 \, + \, 
{T\over{{\rm 273}^{\circ}\mbox{\rm\/C}}}}
\nonumber\\
\nonumber\\
	& = & {\rm\/344.37\ m/s}
\nonumber\\
\nonumber\\
	& \approx & {\rm\/344\ m/s}\ .
\end{eqnarray}

Therefore,
$v_{\mbox{accepted}}$ = 344 m/s.  And thus, we {\it\/can} confirm
the accepted value since it falls within the range of our
measured values~\cite{comparewithstats}.  The percent error in 
this experiment would be
1.8439\% $\approx$ 2\%~\cite{percenterror}. 

\begin{acknowledgments}
The author wishes to thank R. Bichler, J. Bollinger, J. Clawson, D. Mead,
D. Rother,
E. Sivertson, and R. Thompson for participating in PHYS 690: Theory and 
	Practice of Physics Experiments\cite{phys690ss22}, a 
	summer 2022 graduate course at St. 
Cloud State University.  These individuals asked questions that led to 
this investigation.  This work was a direct outgrowth of the summer
	course. 
\end{acknowledgments}


\begin{thebibliography}{99}
        \bibitem{pollard}B. Pollard, R. Hobbs, R. Henderson, M. 
		D. Caballero, and H.J. Lewandowski, Phys. Rev. Phys. Educ.
		Res., {\bf 17}, 010133 (2021).
	\bibitem{wan}T. Wan, Int. Jour. Sci. Educ., 
		http://dio.org/10.1010/09500693.2022.2156824.
	\bibitem{wibig}T. Wibig and P. Dam-o, Phys. Educ., {\bf 40}, 159 (2013).
	\bibitem{taylor}John R. Taylor, {\it\/An Introduction to 
		Error Analysis: The study of Uncertainties in Physical
		Measurements}, Third Edition, University Science Books,
		New York, pg. 190, (2022).
	\bibitem{linest}The function in Microsoft Excel called linest
		elucidates the relationship between the data set and
		the trendline's {\it\/t}-statistics.
		For a discussion of the statistics, see  
J.E. Freund and R.F. Walpole, {\it\/Mathematical 
	Statistics}, 
	Third Edition, Prentice-Hall Inc., pg 430.
	\bibitem{dogma}Laboratory dogma sometimes includes
an expression ``Measurement wins
over theory."  I propose to extend this to ``Measurement wins over
theory and uncertainty wins over statistics''.   Perhaps it is more
useful to express the notion that measurement uncertainty is more 
meaningfully connected to a physical procedure than are 
the mathematical/statistical metrics of standard deviation.  And yet, 
		the two are
not entirely mutually exclusive.
	\bibitem{swokowski}E.W. Swokowski, {\it\/Calculus}, 
%		With Analytic Geometry, 
		2nd Edition, Prindle, Weber \&
		Schmidt, pg 781.
\bibitem{commentoneverything}If possible, known and
		quantifiable effects should be taken into account
		at the point of measurement.  As an example,
		suppose a tape measure is stretched so it will always
		read ``low" by a fixed percent---if the stretch is uniform
		over the entire length of the tape.  That can
		be accounted for most conveniently at the point of 
		measurement. A simple scaling takes care of it.
\bibitem{analysissoftware}Microsoft Excel, Google Sheets, and other
	similar programs are popular choices for carrying out plotting and
	analysis procedures.
\bibitem{matthewswalker}J. Matthews, R.J. Walker, {\it\/Mathematical Methods
	of Physics}, 2nd Edition, The Benjamin/Cummings Publishing Company,
	pg. 387.
\bibitem{richards}M.J. Richards, Phys. Educ., {\bf 6}, 244 (1971).
\bibitem{freundwalpole1}J.E. Freund and R.F. Walpole, {\it\/Mathematical 
	Statistics}, 
	Third Edition, Prentice-Hall Inc., pg 429.
\bibitem{statutility}On the other hand, the standard
deviations in slope and intercept respond more sensitively 
when outliers persist in the data set, while the uncertainties in
the slope and intercept (due to measurement uncertainties) are
more tightly constrained and are therefore less sensitive to
such circumstances.
\bibitem{freundwalpole2}J.E. Freund and R.F. Walpole, {\it\/Mathematical 
	Statistics}, 
	Third Edition, Prentice-Hall Inc., pp 436--438.
\bibitem{s2andsxx}It is commonplace in statistics to utilize
	variance
	\begin{eqnarray}
		\sigma^2 & = & \left({1\over\/N-2}\right)\sum_{i-1}^{N}
		\left[\/y_{i} - (\alpha\/x_{i} + \beta)\right]^{2}\ ,
	\end{eqnarray}
	
	and the correlation metric
	\begin{eqnarray}
		S_{xx} & = & N\,\left[\langle\/x^{2}\rangle - 
		\langle\/x\/\rangle^{2}\,\right]\ .
	\end{eqnarray}
	We adopt those shorthand notations in this work as
		well for brevity and comparison.
\bibitem{klhsoundlab}K.L. Haglin Physics 232 Laboratory Manual, 
	Summer 2022, St. Cloud State University, unpublished.
\bibitem{youngfreedman}H.D. Young and R.A. Freedman, {\it\/University Physics},
	14th Edition, Pearson Publishing, pg. 513
\bibitem{serwayvuille}R.A. Serway, C. Vuille and J. Hughes, {\it\/College
	Physics}, 11th Edition, Cengage Learning, pg. 460.
\bibitem{phys690ss22}This was a professional development opportunity 
	specifically designed
	for laboratory development and discussion.  Purpose, process,
		procedures and reporting expectations for physics
		experiments were reviewed and then undertaken within 
		multiple laboratory activities.  The primary goal was to
		prepare educators to offer newly developed 
		experiments at their home institutions to their own 
		students during upcoming teaching cycles. 
		Specific focus was on the topic of uncertainty.
\bibitem{comparewithstats}We note that if one used the standard deviation
of the slope to formulate a scientific conclusion, the opposite conclusion
would be reached.   This example illustrates the importance of consistently
expressing the uncertainty prior to drawing a scientific conclusion.
\bibitem{percenterror}Since the measured value and the accepted 
		value are reportable to three significant figures, the 
		percent error cannot be expressed past the ones decimal 
		place, meaning
		the tenths place is not reportable.  The reason for this 
		comes from the rules for significant figures when addition
		or subtraction is invoked.  Hence, the percent error must 
		be reported as 2\%. 
\end{thebibliography}
\end{document}